\documentclass[fleqn,10pt]{wlscirep}
\usepackage[utf8]{inputenc}
\usepackage[T1]{fontenc}
\usepackage{textgreek}
\usepackage{multirow}
\usepackage{graphicx}
\usepackage[utf8]{inputenc}
\usepackage{amsmath}
\usepackage{caption}
\usepackage{colortbl}
\title{MoFormer: Multi-objective Antimicrobial Peptide Generation Based on Conditional Transformer Joint Multi-modal Fusion Descriptor}

\author[1]{Li~Wang}
\author[2]{Xiangzheng~Fu}
\author[3]{Jiahao~Yang}
\author[3]{Xinyi~Zhang}
\author[1,*]{Xiucai~Ye}
\author[3,*]{Yiping~Liu}
\author[1]{Tetsuya Sakurai}
\author[3]{Xiangxiang~Zeng}
\affil[1]{Department of Computer Science, University of Tsukuba, Tsukuba 3058577, Japan}
\affil[2]{School of Chinese Medicine, Hong Kong Baptist University, Hong Kong}
\affil[3]{The College of Computer Scienceand Electronic Engineering, Hunan University,
Changsha, Hunan, China}

\affil[*]{Xiucai~Ye(yexiucai@cs.tsukuba.ac.jp), Yiping~Liu(yiping0liu@gmail.com)}


\begin{abstract}
Deep learning holds a big promise for optimizing existing peptides with more desirable properties, a critical step towards accelerating new drug discovery. Despite the recent emergence of several optimized Antimicrobial peptides(AMP) generation methods, multi-objective optimizations remain still quite challenging for the idealism-realism tradeoff. Here, we establish a multi-objective AMP synthesis pipeline (MoFormer) for the simultaneous optimization of multi-attributes of AMPs. MoFormer improves the desired attributes of AMP sequences in a highly structured latent space, guided by conditional constraints and fine-grained multi-descriptor.We show that MoFormer outperforms existing methods in the generation task of enhanced antimicrobial activity and minimal hemolysis. We also utilize a Pareto-based non-dominated sorting algorithm and proxies based on large model fine-tuning to hierarchically rank the candidates. We demonstrate substantial property improvement using MoFormer from two perspectives: (1) employing molecular simulations and scoring interactions among amino acids to decipher the structure and functionality of AMPs; (2) visualizing latent space to examine the qualities and distribution features, verifying an effective means to facilitate multi-objective optimization AMPs with design constraints.
\end{abstract}
\begin{document}

\flushbottom
\maketitle
%
%
\thispagestyle{empty}

\section*{Introduction}

Against the backdrop of antibiotics development significantly trailing behind market demands, Antimicrobial peptides (AMPs) are emerging as promising agents against antimicrobial resistance, offering a compelling alternative to conventional antibiotics\cite{dimasi2016innovation,desselle2017institutional,magana2020value}. Conventional development of AMPs is, however, struggling in dealing with the huge peptide sequence space. This situation motivated a series of studies of deep learning paradigms for AMP discovery\cite{melo2021accelerating,chen2022synthetic,dimasi2016innovation,nagarajan2018computational,nagarajan2019omega76,porto2018joker,porto2018silico}. Early-stage “direct” design approaches, in which a set of AMPs are selected based on expert intuition and screened for a given attribute, are often time-consuming and require extensive resources to explore a small, local region of peptide phase space\cite{jenssen2008qsar,
vishnepolsky2019novo,meher2017predicting,witten2019deep,xiao2013iamp,veltri2018deep,cai2021itp}. Subsequently, “discriminator-guided” approaches emerged, which structures are derived based on models likelihood to exhibit a given attribute value, are desirable as they are far less limited in scope and allow for high-throughput screening of innumerable candidates\cite{murakami2023design,jain2022biological,szymczak2023discovering,hoffman2022optimizing}. In this approach, known as generative deep learning, models learn the natural AMPs sequence landscape in training sets to propose new-to-nature candidates thereby increasing the speed of AMPs discovery.

A variety of deep generative model architectures have been explored for this purpose, with particular focus given to fundamental models, such as the variational autoencoder (VAE)\cite{wang2023accelerating,pandi2023cell,ghorbani2022deep,renaud2023latent,hoffman2022optimizing,dean2021pepvae,tucs2023quantum,szymczak2023discovering}, generative adversarial networks (GANs)\cite{ferrell2020generative,van2021ampgan,cao2023designing,surana2023pandoragan,tucs2020generating}, recurrent neural network (RNN)\cite{capecchi2021machine,muller2018recurrent} and graph neural network (GNN)\cite{buehler2023generative}. Muller et al.\cite{muller2018recurrent} introduced RNNs with long short-term memory (LSTM) units on pattern recognition of helical AMPs, which focused on linear cationic peptides forming amphipathic helices. This model was expected to capture the fundamental feature of amphipathicity closely associated with activity. Szymczak et al. \cite{szymczak2023discovering} proposed HydrAMP, a conditional VAE that learns lower-dimensional, continuous representation of peptides and generated highly active analogs of a peptide with grand antimicrobial activity. Zhang et al.\cite{tucs2020generating} adapted specialized PepGAN architecture with taking the balance between covering active peptides and dodging nonactive peptides.Liu et al.\cite{liu2023evolutionary} introduced an evolutionary multi-objective algorithm that harnessed de novo AMP generation, with antimicrobial activity and diversity as the focal optimization objectives.These efforts collectively highlight the diverse approaches within computational biology to model and enhance our understanding and creation of effective antimicrobial peptides, showcasing the intersection of deep learning with biotechnological advancement.

Despite these encouraging findings, there are still idealism-realism limitations to tackle in the process of producing new AMP with targeted properties. (1) First, discrete sequence data generation, which is more challenging than continuous data generation. Only the singular AMP representation, like amino acid sequence, is difficult to explore the direct structural information and peptide properties that are crucial for generation analysis. (2) Subsequently, it is evident that the  inherent conflict in AMP synthesis are often underestimated, contributing to the high attrition rate of candidates. To cope with the challenging conflict-ridden tasks,  a multi-objective optimization model needs to be introduced with the aim of finding the desired regions in the latent space.

The rapid technological progression within the field of natural language processing may offer some hints towards a future where AI-designed peptides are the norm rather than the exception, with the transformer architecture already becoming the foundation for a multitude of state-of-the-art models, such as BERT\cite{devlin2018bert}, molGPT\cite{bagal2021molgpt}, and their variants. Intuitively, transformer can exhibit very promising results in AMPs synthesis. On one hand, the self-attention mechanism in the Transformer architecture operates on the representation sequence, granting access to any part of it at any given moment\cite{vaswani2017attention}, which is particularly beneficial for comprehending the functional features of AMPs, as these features may be formed by amino acid residues located far from each other in the sequence representation,and the transformer’s attention-based encoding scheme provides for interpretability by analysis of learned attention weights. On the other hand, the latent space of Transformer as a generator is typically characterized by its continuous interpolatability. Through explicit navigation of the sequence space using specific conditions, it enables directed interpolation or sampling, enhancing the model's flexibility and efficiency in generating the desired AMPs.

Given the inherent attributes of AMPs, which result in an intrinsic conflict among their attributes, the simultaneous maximization of all objectives is unattainable. One has to balance them. The best tradeoffs among the objectives can be defined in terms of Pareto optimality. The task of AMPs design can be defined as multi-objective optimization (MOP), an approximation to the Pareto front (PF)\cite{zhang2007moea} is required by a decision maker for selecting a final preferred solution (candidate AMPs). Most MOP algorithms may have many or even infinite Pareto optimal vectors, acquiring a complete PF is often a labor-intensive process, bordering on the unfeasible. Therefore, many MOP algorithms are to find a manageable number of Pareto optimal vectors which are evenly distributed along the PF, and thus good representatives of the entire PF. These good representatives can screen desired AMPs with multifaceted attributes. 

\begin{figure}[ht]
\centering
\includegraphics[width=\linewidth]{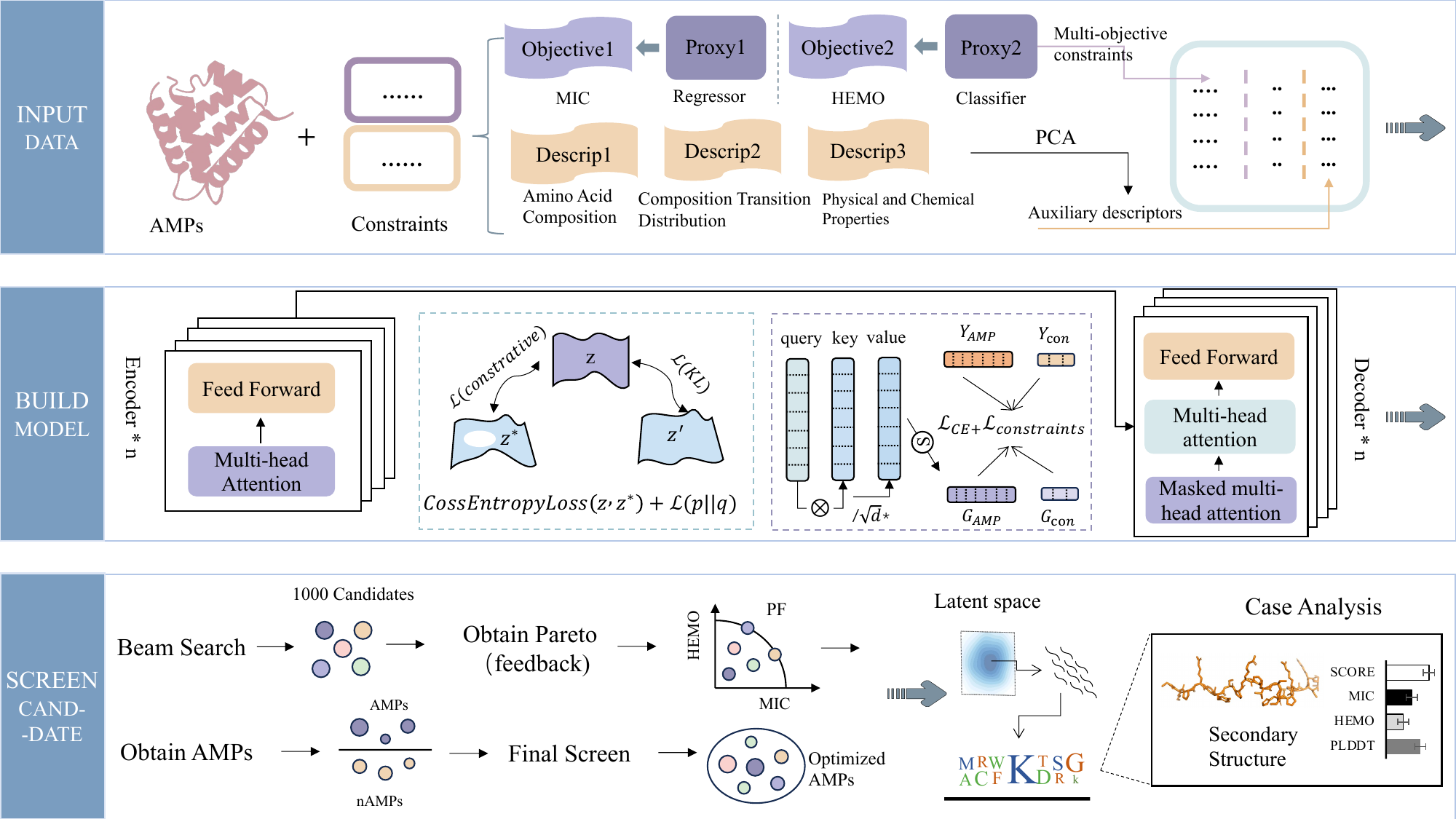}
\caption{The pipeline of MoFormer. We formulated the whole process as three distinct phases: (1) To construct multi-sense conditional constraints, we employ proxies on one hand to ascertain values across multiple objectives, and on the other, we deploy three types of feature descriptors serving as fine-grained auxiliary information; (2) Leveraging the Transformer architecture as our underlying framework, we introduce contrastive regularization techniques and intuitive functional function reshaping of the loss function to optimize the objective space, thereby stably encouraging the generation of desired AMPs; (3) We develop a dependable and swift screening protocol that employs a hierarchical filtering system based on Pareto non-dominated sorting and proxies fine-tuned from large models.Subsequently, molecular simulations and in-depth interpretative analyses were executed on the selectively isolated AMPs.}
\label{framework}
\end{figure}

In this study, we propose MoFormer, a deep transformer-based approach to AMPs design, which leverages conditional constraints and fine-grained descriptors to produce desired and deliberate candidates in regularized latent space. The overall framework of the proposed approach is shown in Figure 1\ref{framework}. The contributions are generalized as follows:

(1) To synthesis tailor-made and desired AMPs, we establish a multi-objective attributes and multi-fused descriptors condition Transformer (MoFormer) for optimizing the generation progress, multi-objective attributes from proxy computations as tailored guidance, multi-fused descriptors for enhancing the diversity and accuracy of AMPs. Results show that MoFormer, enhanced with fine-grained descriptors, performs exceptionally well, surpassing MoFormer without descriptors (MoFormer w/o D) across various multi-objective scenarios.

(2) To enhance the performance of MoFormer, we implement strategies on two fronts: on one hand, diminishing the weight of the KL divergence regularization term and incorporating a contrastive regularization item to prevent mode collapse; on the other hand, by employing intuitive design techniques based on functional functions to guide the synthesis of AMPs characterized by high antimicrobial activity and minimal hemolytic effects.

(3)To streamline the selection process of candidates, we integrated a fast non-dominated sorting algorithm alongside a suite of fine-tuned proxies derived from large models, facilitating a hierarchical evaluation. This methodological blend proved instrumental in pinpointing 56 AMP candidates, distinguished by their bespoke multi-objective attributes of enhanced antimicrobial efficacy and reduced hemolytic potential.

(4) Empirical Validation of MoFormer: we engineered the MoFormer triplet model employing a strategic ablation study and engaged in comparative analyses against state-of-the-art methodologies.The findings underscore our model's superior efficacy in synthesizing targeted AMPs. Leveraging molecular simulations and latent space visualization techniques, we endeavor to elucidate the intricate structure and functionality of these novel AMPs.

\section*{Results}

\subsection*{MOFormer Outperforms Other Models}
Moformer designed for the generation of AMPs with Multiple attributes, while satisfying specified Optimization constraints. Considering the significant correlation among antibacterial activity (MIC), hemolysis (HEMO), and AMP sequences, we employed these elements as multi-objective constraints, and further leverage fine-grained descriptors guiding the generative process. MoFormer is trained in three modes: MoFomer(w/o D), MoFormer(w/o F), MoFormer. In MoFomer(w/o D) generation mode, the model is trained to generate AMPs single depend on multiple attributes constraints. Training in the MoFormer(w/o F) model facilitates the model to fine-grained capture AMP sequences distinctiveness apart from multiple attributes constraints. This is achieved by including sequence, structure and physicochemical descriptors that have proved significantly contributed to AMP biological function. Finally, during the MoFomer generation mode, the model combined a multi-objective optimization algorithm with a feedback loop to condition and optimize (fine-tune) the generation process.

The performance of our proposed methods was compared to with six state-of-the-art methods: LSTM\cite{muller2018recurrent}, AMP\_GAN\cite{lin2021discovering}, PepGAN\cite{tucs2020generating}, WAE\cite{lin2021discovering}, AMPEMO\cite{liu2023evolutionary} and HMAMP. LSTM method trained RNNs with LSTM units on pattern recognition of helical AMPs and used the resulting model for de novo sequence generation. AMP-GAN method trained a deep convolutional GAN with known AMPs to generate high antimicrobial activity candidates. PepGAN method controlled the probability distribution of generated sequences to cover active peptides as much as possible. WAE method combined the wasserstein autoencoder with a particle swarm optimization forward search algorithm to screen anticancer peptides with desired attributes. We adapted this approach for AMP datasets to assess the generation performance of the encoder. These methods are controlled to yield AMPs that align with the single attribute. AMPEMO is a multi-objective evolutionary method designed to optimize both the antimicrobial activity and diversity of AMPs. HMAMP introduced a novel approach based on GAN, hypervolume-driven multi-objective AMP generation, for the de novo design of AMPs with desired attributes. For AMPEMO and HMAMP, despites both have made notable advances in the realm of multi-objective optimization, the exploration of latent space optimization techniques, predicated on multi-objective constraints, remains an untapped area of research.

\begin{figure}[p]
\centering
\includegraphics[width=\linewidth]{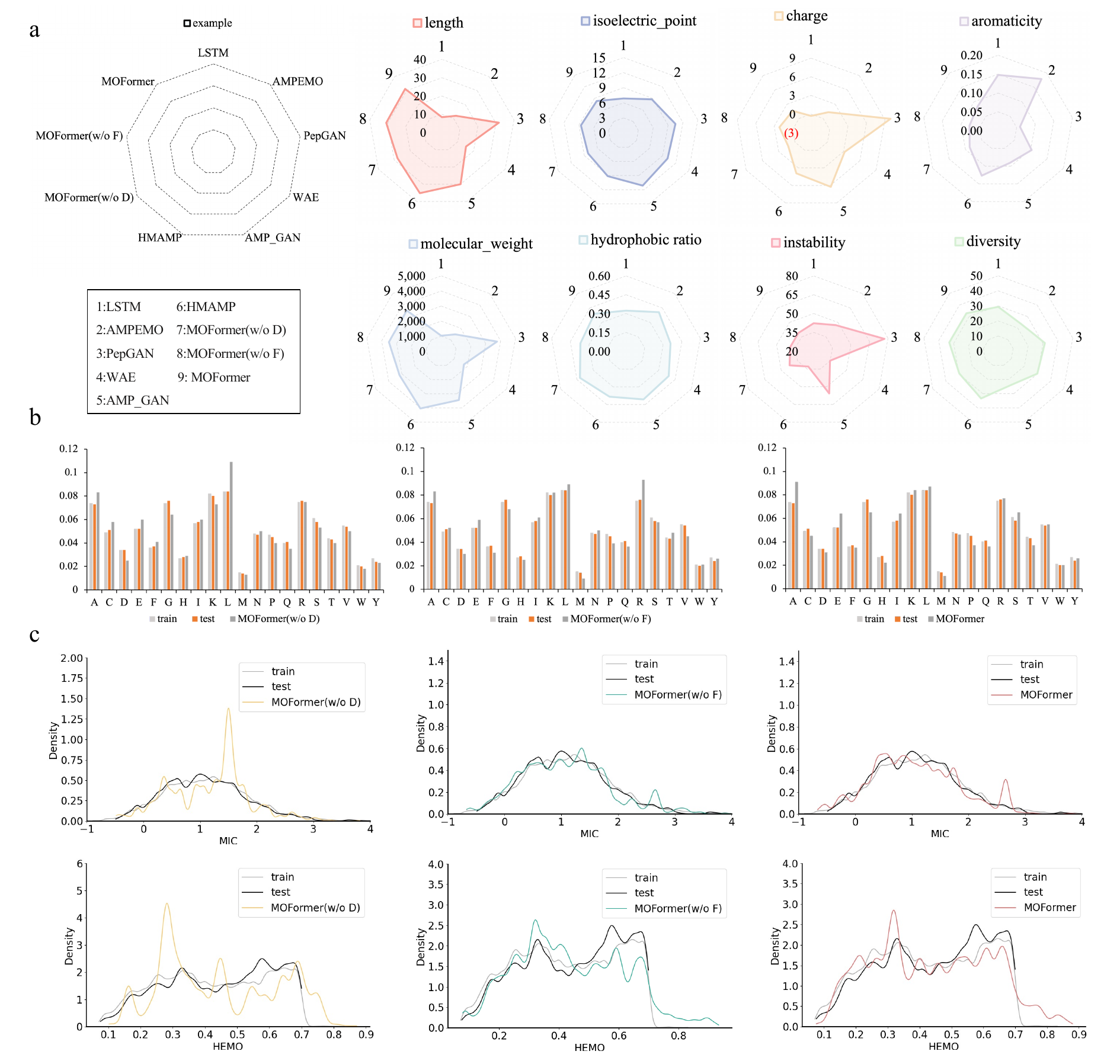}
\caption{a.The comparison of the fundamental properties of AMPs generated by MOFormer and other state-of-the-art methods. b.Comparison of amino acid distribution between AMP generated by MoFormer, MoFomer(w/o D), MoFomer(w/o F)datasets. c.Comparison of attribution distribution between AMP generated by MOFormer, MoFomer(w/o D), MoFomer(w/o F) and benchmark dataset. }
\label{basic_fig}
\end{figure}

\begin{table}[ht]
\centering
\begin{tabular}{|l|l|l|l|l|}
\hline
method          & MIC(\textless{}2) & HEMO(\textless{}0.5) & Combination & HV    \\ \hline
LSTM            & 0.63              & 0.831                & 0.498       & 0.566 \\ \hline
AMPEMO          & 0.82              & 0.443                & 0.347       & 0.439 \\ \hline
PepGAN          & 0.152             & 0.945                & 0.143       & 0.718 \\ \hline
WAE             & 0.841             & 0.338                & 0.245       & 0.591 \\ \hline
AMP\_GAN        & 0.984             & 0.393                & 0.381       & 0.72  \\ \hline
HMAMP           & 0.94              & 0.35                 & 0.305       & 0.891 \\ \hline
MOFormer(w/o D) & 0.622             & 0.898                & 0.575       & 0.767 \\ \hline
MOFormer(w/o F) & 0.622             & 0.886                & 0.518       & 0.93  \\ \hline
MOFormer        & 0.596             & 0.881                & 0.502       & 0.976 \\ \hline
\end{tabular}
\caption{The comparison of the interval satisfaction proportions of AMPs generated by MoFomer versus other methods, along with a comparison of hypervolume multi-objective metric, aimed toward antimicrobial activity(MIC) and hemolytic(HEMO) properties.}
\label{MH_tab}
\end{table}

Let us illustrate or findings using antimicrobial activity(MIC) and hemolysis(HEMO) as an example. We first trained and sampled 1000 AMPs from each model, and then applied characteristic-related metrics and the multi-objective evaluation metric hypervolume (HV) for a comprehensive evaluation. In detail, we consider characteristic-related metrics including length, isoelectric point, charge, aromaticity, molecular weight, instability, hydrophobic ratio and diversity as clearly shown in Figure \ref{basic_fig}(a). AMP typically range in length from 10 to 60 amino acids, Generally, shorter peptides are easier to synthesize and may penetrate bacterial membranes more readily but can also increase hemolysis due to their smaller size. AMPs with a higher isoelectric point become positively charged in the body's pH, aiding their attachment to negatively charged bacterial membranes, enhancing their antimicrobial effects. Yet, a very high isoelectric point can raise the risk of damaging human cells due to increased hemolytic activity. The positive charge of AMP boosts their antimicrobial action by binding to bacterial membranes, but too much charge can harm human cells by increasing hemolysis. Aromatic amino acids can insert into membranes, promoting antimicrobial action. Larger AMPs, due to their molecular weight, may penetrate less but offer higher specificity and reduced risk of damaging human cells. A balanced hydrophobicity in AMPs improves membrane interaction, but too much can disrupt cells non-specifically and increase hemolysis. The instability of AMPs refers to their rate of degradation within the organism. Unstable peptides might require special delivery systems or chemical modifications to increase their half-life in the body, allowing them to be more effective. The characteristics of AMP directly influence their ability to combat microbes and their potential to harm host cells. Ideally, these peptides should effectively destroy bacteria with minimal risk to the host's cells, balancing potent antimicrobial properties against low hemolytic effects. Upon further investigation into the diversity of AMP generation, we found that MOFormer, although not the top performer, ranks well within the higher tier of comparative methods in terms of its efficacy.

\begin{figure}[h]
\centering
\includegraphics[width=\linewidth]{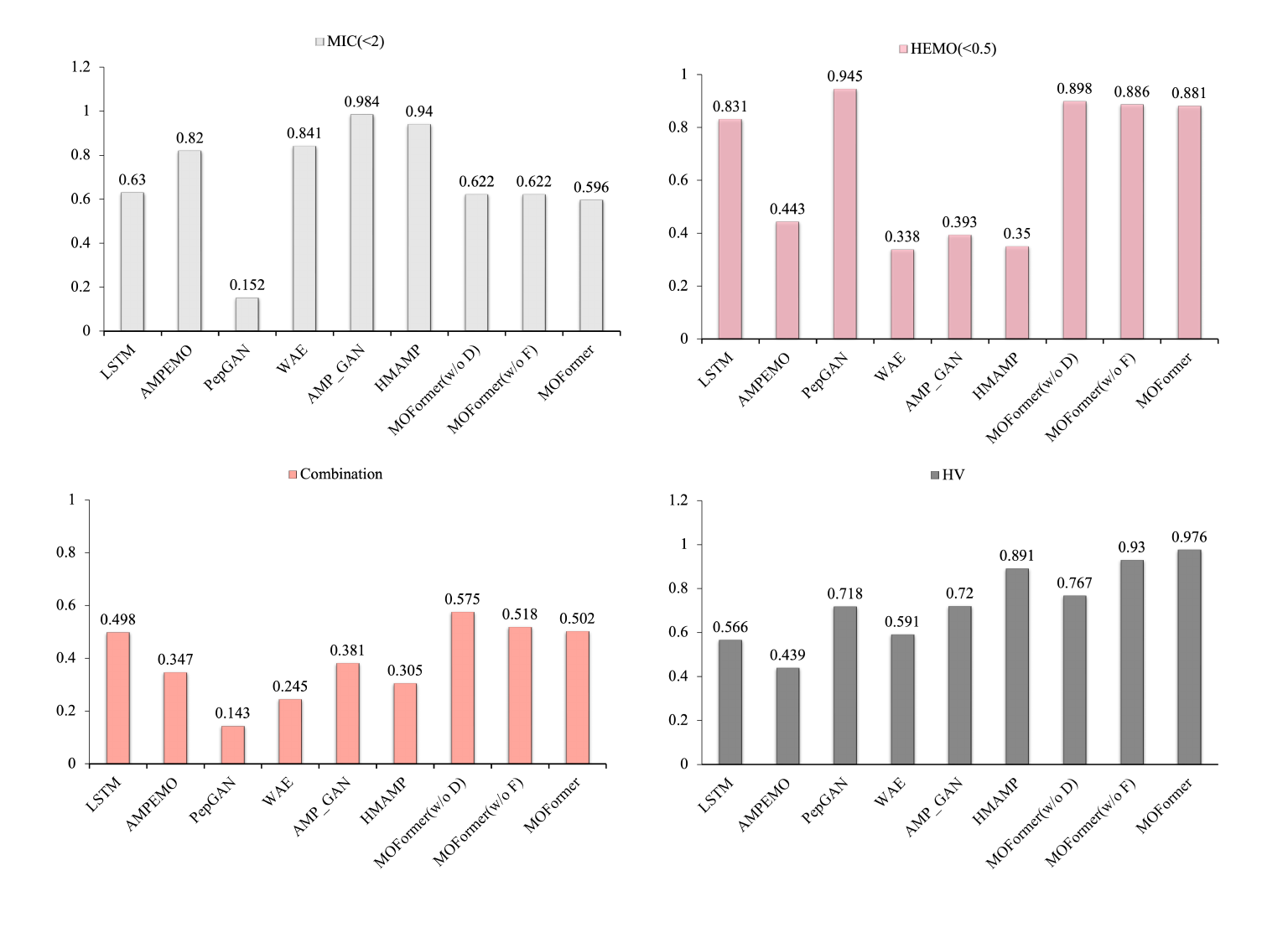}
\caption{The distribution of the interval satisfaction proportions of AMPs generated by MoFomer versus other methods, along with a comparison of HV multi-objective metric, aimed toward antimicrobial activity(MIC) and hemolytic(HEMO) properties.}
\label{compare_fig}
\end{figure} 

The HV, also known as S metric, is an unary metric that measures the size of the objective space covered by an approximation set. A reference point $\eta$  must be used to calculate the mentioned covered space. In this respect, we set the $\eta$ as (2,0.5). Intuitively, we expect the log(MIC) regression values for AMPs to be less than 2, with the probability of hemolytic activity being below 0.5. HV considers all three aspects: accuracy, diversity and cardinality and has been widely accepted in multi-objective scenario. For more intuitive comparison, Table \ref{MH_tab} and Figure\ref{compare_fig} summarizes the performance of the generation process, showing the percentage of peptides falling within each attribute (MIC and Hemolytic) range, the percentage of candidate AMPs meeting all attribute criteria, as well as HV metric. In terms of MIC(<2) and HEMO(<0.5) columns,  although the AMPs produced by our approach do not achieve the top marks in interval occupancy ratio, their overall performance is noteworthy, especially HEMO. We believe this is due to our method's strategy of balancing multiple objectives, rather than disproportionately favoring any single attribute. For combination of MIC and HEMO, we observe that our methods(MoFomer(w/o D), MoFormer(w/o F), MoFormer) perform the best in terms of combination, MoFormer is slightly inferior to MoFomer(w/o D) and MoFormer(w/o F). This might be attributed to the fact that MoFormer prefer to yield lower MIC value and less HEMO probability, while MoFomer(w/o D), MoFormer(w/o F) generate a larger fraction of candidates approaching the right critical threshold. Notably, out of a total of 9 methods, MoFormer achieves the best HV result focusing on advancing in the opposite direction of the reference point, suggesting MoFormer can control the best the balance between MIC and HEMO. 

\begin{table}[ht]
\centering
\begin{tabular}{|l|l|l|l|l|}
\hline
method          & MIC(\textless{}2) & TOXI(\textless{}0.5) & Combination & HV    \\ \hline
LSTM            & 0.369             & 0.63                 & 0.277       & 0.779 \\ \hline
AMPEMO          & 0.749             & 0.842                & 0.657       & 0.69  \\ \hline
PepGAN          & 0.777             & 0.945                & 0.743       & 1.081 \\ \hline
WAE             & 0.617             & 0.881                & 0.565       & 1.06  \\ \hline
AMP\_GAN        & 0.646             & 0.984                & 0.64        & 0.946 \\ \hline
HMAMP           & 0.722             & 0.956                & 0.706       & 1.154 \\ \hline
MOFormer(w/o F) & 0.731             & 0.923                & 0.696       & 1.138 \\ \hline
MOFormer        & 0.745             & 0.936                & 0.717       & 1.291 \\ \hline
\end{tabular}
\caption{The comparison of the interval satisfaction proportions of AMPs generated by MoFomer versus other methods, along with a comparison of HV multi-objective metric, aimed toward antimicrobial activity(MIC) and toxicity(TOXI).}
\label{MT_tab}
\end{table}

To validate the superiority and universality of our meticulously designed MoFormer, we proceeded to analyze its performance in terms of MIC and TOXI(toxicity) compared to LSTM\cite{muller2018recurrent}, AMP\_GAN\cite{lin2021discovering}, PepGAN\cite{tucs2020generating}, WAE\cite{lin2021discovering}, AMPEMO\cite{liu2023evolutionary} and HMAMP. In assessing both the MIC proportions and TOXI proportions individually(see Table \ref{MT_tab} ), methods MoFormer(w/o F) and MoFormer emerge as frontrunners, particularly method MoFormer, which exhibits a distinct advantage. Hence, it follows that in the realm of multi-attribute combination, MoFormer(w/o F) and MoFormer significantly outperform comparative methods, albeit with a slight underperformance relative to PepGAN. Nevertheless, this minor shortfall does not detract from the innovative nature of our methods. It is widely acknowledged that toxicity data are inherently scarce and challenging to gather, which inherently influences the validation process for methods MoFormer(w/o F) and MoFormer. As a final point, our observations reveal that method MoFormer excels in the HV metric, a result that is in perfect harmony with our expectation of minimizing two-objective within a bidimensional framework,  highlighting the superior and robust nature of our optimization strategy.

\subsection*{Characteristics of The Designed AMPs}
To check the homology of MoFormer-generated AMP sequences with train and test dataset, we compare the distribution of amino acid composition as shown in Figure \ref{basic_fig}(b). MoFomer(w/o D) generated AMP sequences show distinct character: specifically, these sequences are richer in Ala, Cys, Glu and Leu, whereas Asp, Gln, Thr, Val and Trp content is reduced, MoFomer(w/o F) generated AMP sequences show distinct character: specifically, these sequences are richer in Ala, Leu and Arg, whereas Phe, Met, Pro, Gln and Val content is reduced, MoFomer generated AMP sequences show distinct character: specifically, these sequences are richer in Ala, Glu, Leu, Arg and Thr, whereas Asp, Gly, Phe, Met, Gln and Val content is reduced, in comparison to train and test antimicrobial sequences. Hydrophilic amino acids, including Ala, Glu, Ser, Gln, are known for their ability to interact with hydrophobic domains within microbial cell membranes, potentially leading to membrane disruption. On the other hand, hydrophobic amino acids, such as Leu, Ala, and Thr, have the propensity to penetrate microbial membranes and compromise their structural integrity. Positively charged amino acids like Arg and Lys disrupt both microbial and host cell membranes by interacting with negatively charged areas, offering antibacterial benefits but also increasing hemolytic activity.

We also present the distribution of MoFormer-generated MIC and HEMO with train and test dataset as shown in Figure \ref{basic_fig}(c). In terms of MIC, upon observation, it has been determined that MoFomer(w/o F) and MoFomer exhibit a heightened ability to assimilate the distribution of the original dataset when contrasted with MoFomer(w/o D). Notably, MoFomer manifested a consistent leftward trend as anticipated, which signifies an augmentation of the antimicrobial efficacy, an advancement that can be credited to the optimization processes implemented by MoFomer. For HEMO, MoFomer(w/o F) and MoFomer exhibit a distinct trend towards reduced hemolytic activity, with their distributions showing a notable leftward shift when compared to MoFomer. The extent of this optimization appears to be more significant in relation to MIC. Despite this, all proposed methods have produced a small fraction of AMPs with values above 0.8, which could be attributed to the intrinsic trade-offs encountered during multi-objective optimization, complicating the model's ability to equilibrate various properties. Interestingly, we further employ the T-test to conduct a significance analysis of MoFomer and other datasets(train,test, MoFomer(w/o D), and MoFomer(w/o F)), aiding in our understanding of the differences between the samples.As shown in Table \ref{T-test}, MoFomer can optimize lower MIC value than train(test p value < 0.05) and MoFomer(w/o D) (test p value < 0.001). Additionally, MoFomer’s performance shows a significant difference compared to train in terms of HEMO. MoFomer(w/o D), and MoFomer(w/o F) fail to produce AMPs with significantly higher performance than well-designed MoFomer of the considered multiple attribution.

\begin{table}[ht]
\centering
\begin{tabular}{|l|llll|}
\hline
\multicolumn{1}{|c|}{\multirow{2}{*}{T-test}} & \multicolumn{4}{c|}{MOFormer} \\ \cline{2-5} 
\multicolumn{1}{|c|}{}                        & \multicolumn{1}{l|}{train}  & \multicolumn{1}{l|}{test}  & \multicolumn{1}{l|}{MOFormer(w/o D)} & MOFormer(w/o F) \\ \hline
MIC                                           & \multicolumn{1}{l|}{0.0234} & \multicolumn{1}{l|}{0.381} & \multicolumn{1}{l|}{0.0006}          & 0.595           \\ \hline
HEMO                                          & \multicolumn{1}{l|}{0.0462} & \multicolumn{1}{l|}{0.343} & \multicolumn{1}{l|}{0.3307}          & 0.854           \\ \hline
\end{tabular}
\caption{T-test statistics for multi-attribute comparison of AMPs generated by method MoFomer, ablation methods, and the benchmark dataset.}
\label{T-test}
\end{table}

\subsection*{Structural and Mechanistic Analyses}
\begin{figure}[h]
\centering
\includegraphics[width=\linewidth]{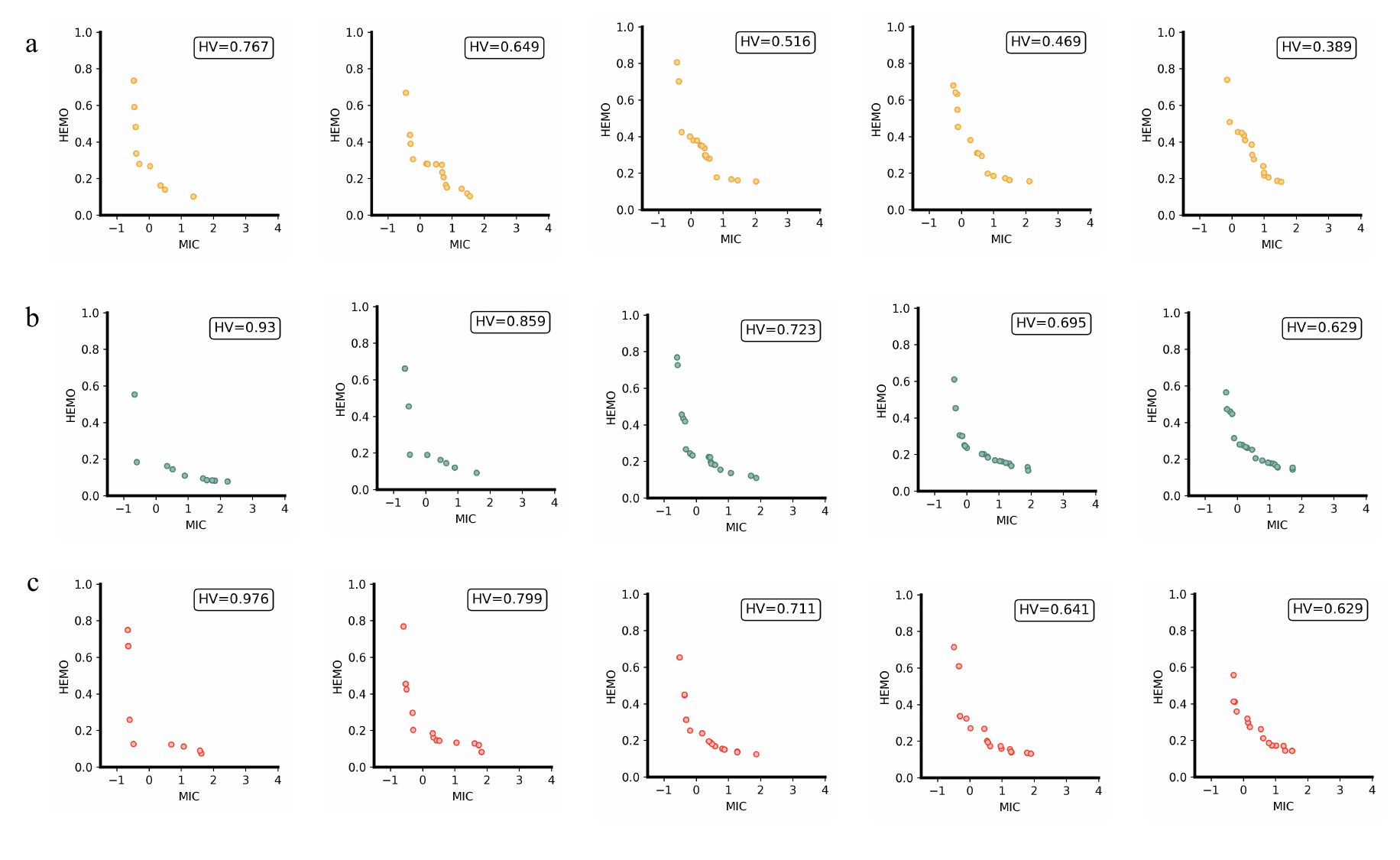}
\caption{a,b and c show the top five PFs of MoFomer(w/o D), MoFomer(w/o F) and MoFomer, respectively.}
\label{pareto_fig}
\end{figure}

To screen the 1000 MoFormer-generated AMP sequence, we first used multi-objective optimization (MOP) algorithm that generate top-five pareto front by non-dominated sorting. Observing Figure \ref{pareto_fig}, we can discern that the two minimal objectives among the top-five pareto front can satisfy our expectations to the greatest extent, achieving high antibacterial activity and low hemolysis. Through the lens of prior experimental evaluations, MoFomer has emerged as the paramount approach for the synthesis of AMP, standing superior to MoFomer(w/o D) and MoFomer(w/o F) amidst constraints posed by multiple objectives. Consequently, from the top-five PFs of MoFomer, we curated a collection of 67 candidates for an in-depth structural and functional scrutiny. Furthermore, these 67 candidates underwent a rigorous sifting process via a conventional AMP classifier SVM, subsequently yielding 56 candidates with high antibacterial activity and low hemolytic potential.

\begin{figure}[p]
\centering
\includegraphics[width=\linewidth]{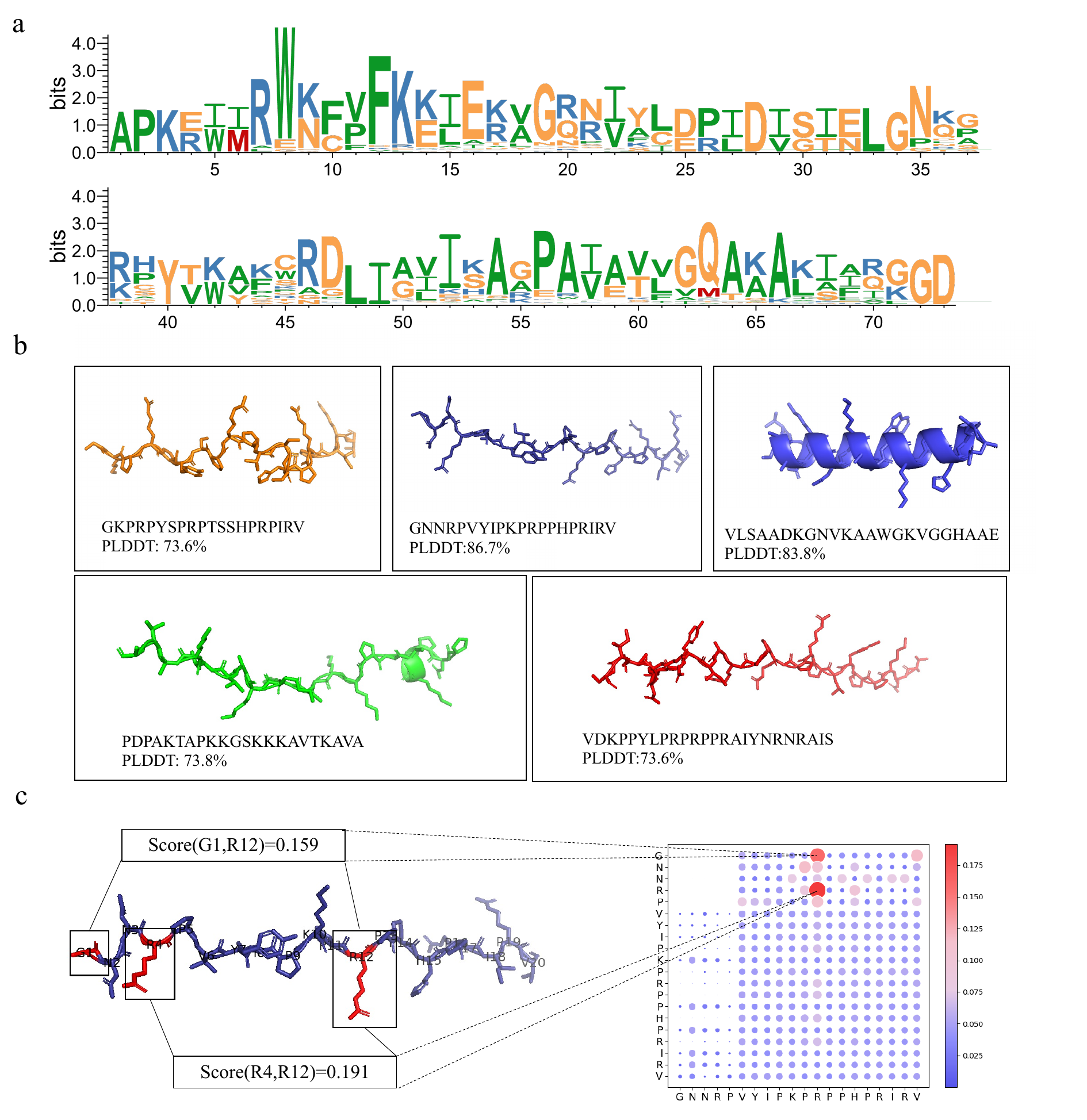}
\caption{Consider antimicrobial activity and hemolytic properties. a.sequence logos of candidate peptides generated by MoFormer. b.AMP structures according to molecular simulations for ID1-ID5. c.Attention weights can be extracted from the transformer module, delving deeper into the interpretation of AMP structure and function.}
\label{case1_fig}
\end{figure} 

To elucidate the functional attributes of these 56 contenders with greater specificity, we employ sequence logos to illustrate the conservation and diversity of amino acids at specific positions within this group of sequences. Within the ambit of AMP sequences, it is the conserved amino acids that frequently harbor functional significance, being integral to enzymatic catalytic cores or mediating interfaces for protein-protein interactions. These conserved residues serve as beacons, pinpointing the loci that are pivotal in modulating AMP efficacy, thereby offering invaluable clues that advance our exploration of AMP sequence functionality and structural conformation. Figure \ref{case1_fig}(a) elucidates regions of pronounced amino acid conservation, such as lysine (K) and arginine (R), which bear positive charges that are integral to the AMP' efficacy. These residues engage in electrostatic interactions with the negatively charged bacterial surfaces, facilitating the adherence and subsequent membrane translocation of the peptides. Figure \ref{case1_fig}(a) also illustrates the prevalence of hydrophobic and polar residues like phenylalanine (F), leucine (L), glycine (G), and serine (S). Hydrophobic residues play a pivotal role in the interaction with bacterial membranes, whereas polar residues modulate the solubility and in vivo stability of the peptides. Generally, an excess of highly hydrophobic or positively charged amino acid residues might increase hemolytic potential. Despite the presence of conserved positions, the sequence diversity remains appreciable—a testament to the nuanced balance of amino acid composition that mitigates hemolytic risk while preserving antimicrobial potency.

\begin{table}[ht]
\centering
\resizebox{\textwidth}{!}{
\begin{tabular}{|l|l|l|l|l|l|l|l|l|l|l|l|}
\hline
ID & sequence                 & length & isoelectric\_point & charge & aromaticity & molecular\_weight & hydrophobic ratio & instability & MIC(log)                      & HEMO                         & TOXI                         \\ \hline
1  & GKPRPYSPRPTSSHPRPIRV     & 20     & 11.999             & 4.588  & 0.05        & 2285.609          & 0.1               & 49.225      & {\color[HTML]{080808} 0.184}  & {\color[HTML]{080808} 0.24}  & -                            \\ \hline
2  & GNNRPVYIPKPRPPHPRIRV     & 20     & 11.999             & 4.588  & 0.05        & 2363.767          & 0.2               & 25.815      & {\color[HTML]{080808} -0.364} & {\color[HTML]{080808} 0.447} & -                            \\ \hline
3  & VLSAADKGNVKAAWGKVGGHAAE  & 23     & 8.477              & 0.555  & 0.043       & 2236.485          & 0.434             & -16.669     & {\color[HTML]{080808} 1.24}   & {\color[HTML]{080808} 0.157} & -                            \\ \hline
4  & PDPAKTAPKKGSKKKAVTKAVA   & 22     & 10.4               & 5.884  & 0           & 2221.642          & 0.318             & 15.909      & {\color[HTML]{080808} 0.557}  & {\color[HTML]{080808} 0.195} & -                            \\ \hline
5  & VDKPPYLPRPRPPRAIYNRNRAIS & 24     & 11.451             & 4.515  & 0.083       & 2847.283          & 0.25              & 32.483      & {\color[HTML]{080808} 0.14}   & {\color[HTML]{080808} 0.298} & -                            \\ \hline
6  & KWKFKKIPKFLHLAKKF        & 17     & 10.778             & 6.576  & 0.235       & 2187.757          & 0.411             & -6.117      & {\color[HTML]{080808} 0.246}  & {\color[HTML]{080808} -}     & {\color[HTML]{080808} 0.051} \\ \hline
7  & YCYCRRRFCVCVGR           & 14     & 9.303              & 3.454  & 0.214       & 1784.163          & 0.5               & 74.278      & {\color[HTML]{080808} -0.702} & {\color[HTML]{080808} -}     & {\color[HTML]{080808} 0.272} \\ \hline
8  & RGGRLCYCRRRFCVCT         & 16     & 9.88               & 4.456  & 0.125       & 1949.357          & 0.437             & 99.043      & {\color[HTML]{080808} -0.452} & {\color[HTML]{080808} -}     & {\color[HTML]{080808} 0.283} \\ \hline
9  & KWKKFKKKLAGLLAKVLTT      & 19     & 10.778             & 6.539  & 0.105       & 2201.781          & 0.421             & -7.37       & {\color[HTML]{080808} 0.044}  & {\color[HTML]{080808} -}     & {\color[HTML]{080808} 0.116} \\ \hline
10 & KWARLWRWFRITRWLWYIK      & 19     & 11.999             & 5.549  & 0.368       & 2765.311          & 0.315             & 88.81       & {\color[HTML]{080808} 0.863}  & {\color[HTML]{080808} -}     & {\color[HTML]{080808} 0.11}  \\ \hline
11 & KKKTWWKTWTKWSQPKK        & 17     & 10.778             & 6.539  & 0.235       & 2275.694          & 0                 & 5.547       & {\color[HTML]{080808} -0.062} & {\color[HTML]{080808} -}     & {\color[HTML]{080808} 0.134} \\ \hline
\end{tabular}
}
\caption{Essential properties and targeted attribute values of the meticulously curated ID1-ID11 AMPs.}
\label{case_tab}
\end{table}

In the realm of antimicrobial research, short-chain AMPs stand out due to their compact molecular architecture and straightforward structural configurations, which confer an enhanced ability to traverse bacterial membranes and biofilms. Moreover, their simplicity renders them less susceptible to enzymatic degradation, thus bolstering their in vivo stability and therapeutic potency. With the optimization goal of maximizing antimicrobial proficiency while minimizing hemolytic activity, we have selected 5 candidates with a length of 25 or fewer amino acids from the aforementioned 56 candidates for molecular simulation and functional analysis. For these select candidates, labeled ID1-ID5, we have executed structural predictions employing the cutting-edge AlphaFold platform and have proceeded with molecular simulations via the PyMOL software suite. In Figure \ref{case1_fig}(b) and Table \ref{case_tab},ID1, ID2, ID4, and ID5 display protein structures that are primarily composed of random coil structures. Such a conformational trait imparts a significant degree of malleability, enabling these peptides to dynamically interface with bacterial membranes, accommodating a diverse array of lipid compositions and structural characteristics. The absence of a stable, fixed architecture may confer an advantage, streamlining the interaction process with bacterial membranes, thus expediting the antimicrobial assault. Furthermore, the PLDDT scores spanning from 73 to 87\% instill a measured confidence in the predictive accuracy of these structural models. Notably, peptides with higher PLDDT scores are indicative of a robust structural definition, approximating the functional form within biological contexts, thereby potentially enhancing the probability of potent antimicrobial efficacy. ID3's structure is delineated by an array of $alpha$-helices, a structural motif commonly associated with efficacious bacterial membrane interactions. The characteristic helical conformation enables these α-helices to integrate into the bacterial lipid bilayer, perturbing membrane stability and culminating in bacterial cytotoxicity. The robust PLDDT score of 83.8\% augurs a high degree of structural fidelity in vivo, which could translate to considerable resistance against enzymatic degradation and an extended circulatory persistence, fortifying the antimicrobial peptide's potential as a durable therapeutic agent.

To explore decipher the intricate interplay between amino acid residues within the amphitropic regions of peptides, we harnessed the attention mechanism embedded within the MoFormer model for a nuanced, quantitative interpretation of AMP sequences(see Figure \ref{case1_fig}(c)). Focusing on the sequence represented by ID2, distinguished by its apex PLDDT score, we noted a pronounced attention score between the residues R1 and R12. This observation intimates a formidable interaction between the duo of arginine residues, potentially orchestrated through a symphony of hydrogen bonds, salt bridges, and other non-covalent modalities, thereby contributing to the conformational fidelity of the AMP sequence. Additionally, the interplay between G1 and R12 garnered significant attention, underscoring the role of the electrostatic interplay between the positively charged arginine and the sterically unencumbered glycine in the structural integrity and folding paradigm of the peptide. Collectively, these insights gleaned from the attention score landscape are imperative for elucidating the AMP's mode of engagement with bacterial membranes and its subsequent antimicrobial action mechanism.

\begin{figure}[p]
\centering
\includegraphics[width=\linewidth]{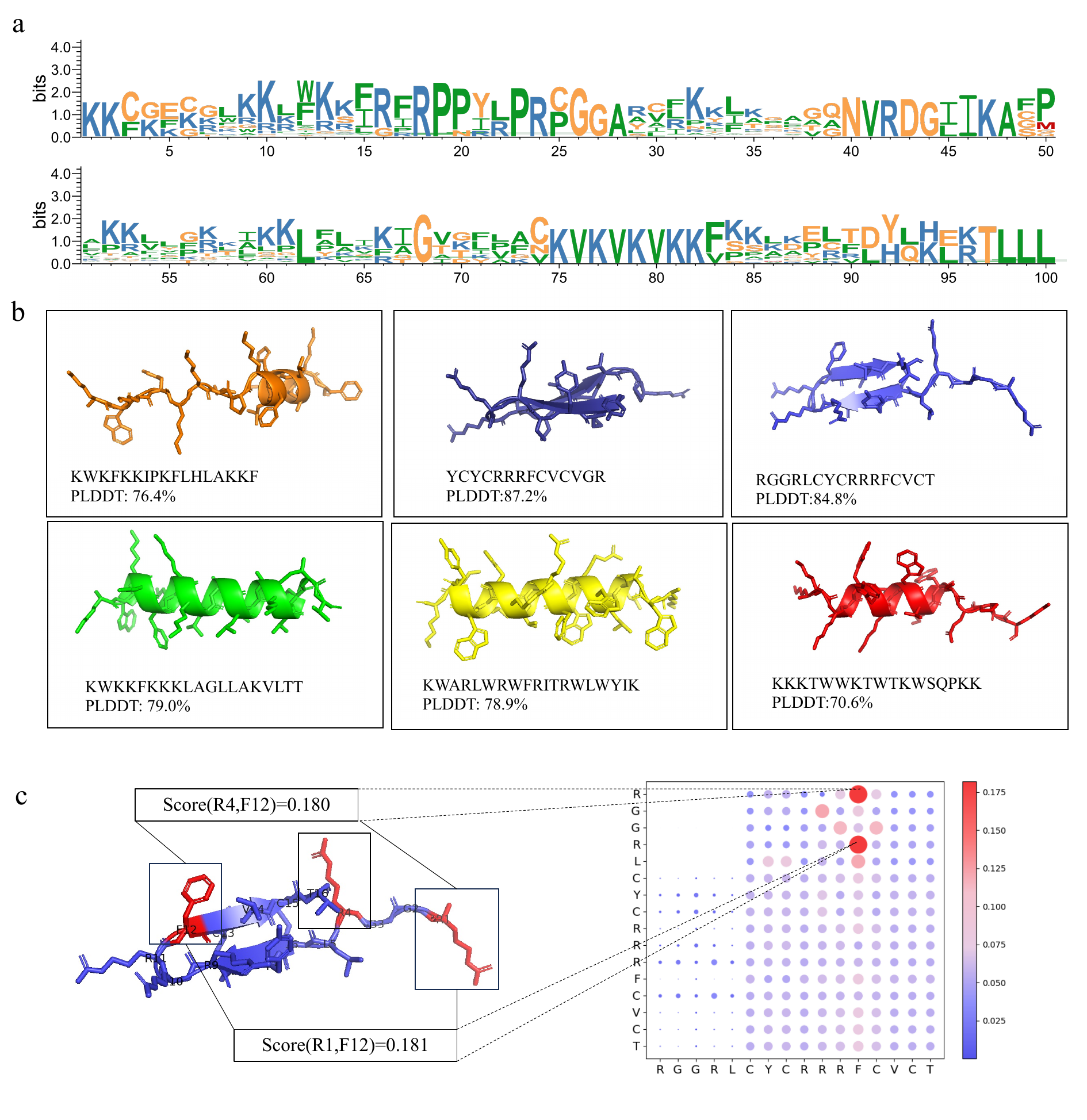}
\caption{Consider antimicrobial activity and toxicity. a.sequence logos of candidate peptides generated by MoFormer. b.AMP structures according to molecular simulations for ID6-ID11. c.Attention weights can be extracted from the transformer module, delving deeper into the interpretation of AMP structure and function.}
\label{case2_fig}
\end{figure} 

To substantiate the viability of MoFormer, we meticulously replicated the screening mechanism and process for high antimicrobial activity and low hemolytic activity, conducting an in-depth analysis of the generated results targeting high antimicrobial efficacy and low toxicity. Initially, we elucidated the sequence logos of 67 candidate peptides as shown in Figure \ref{case2_fig}(a) and Table \ref{case_tab}, highlighting the prevalence of positively charged residues like lysine (K) and arginine (R), which are instrumental in mediating interactions between AMPs and the typically negatively charged bacterial cell membranes. Concurrently, hydrophobic residues such as leucine (L) and phenylalanine (F) are beneficial for engaging with bacterial membranes when appropriately balanced, yet an overabundance might exacerbate the perturbation of cell membranes, thereby augmenting the cytotoxic potential of these peptides. Following this, we refined our pool to six candidates not exceeding 20 amino acids in length, and we proceeded to render their structural sequence visualizations(see Figure \ref{case2_fig}(b)). IDs 6-11 predominantly revealed the presence of $alpha$-helices, with PLDDT scores oscillating between 70-90\%, which correlates with high activity levels and substantial structural prediction reliability. Ultimately, we constructed a detailed heatmap of amino acid interactions for ID8 to quantitatively decipher the inter-residue relationship scores in Figure \ref{case2_fig}(c), thereby shedding light on the sequence's functionality. The arginine (R) residues, with their capacity to form hydrogen bonds, and phenylalanine (F) residues, noted for their hydrophobic character, may contribute significantly to the peptide's conformational stability, particularly when synergistic hydrophobic interactions are present to reinforce their connective integrity. Overall, the gathered interaction metrics are pivotal in unraveling the mechanisms by which proteins consolidate their three-dimensional conformation and profoundly influence their biological efficacy.

\subsection*{Latent Space Visualization}
\begin{figure}[h]
\centering
\includegraphics[width=\linewidth]{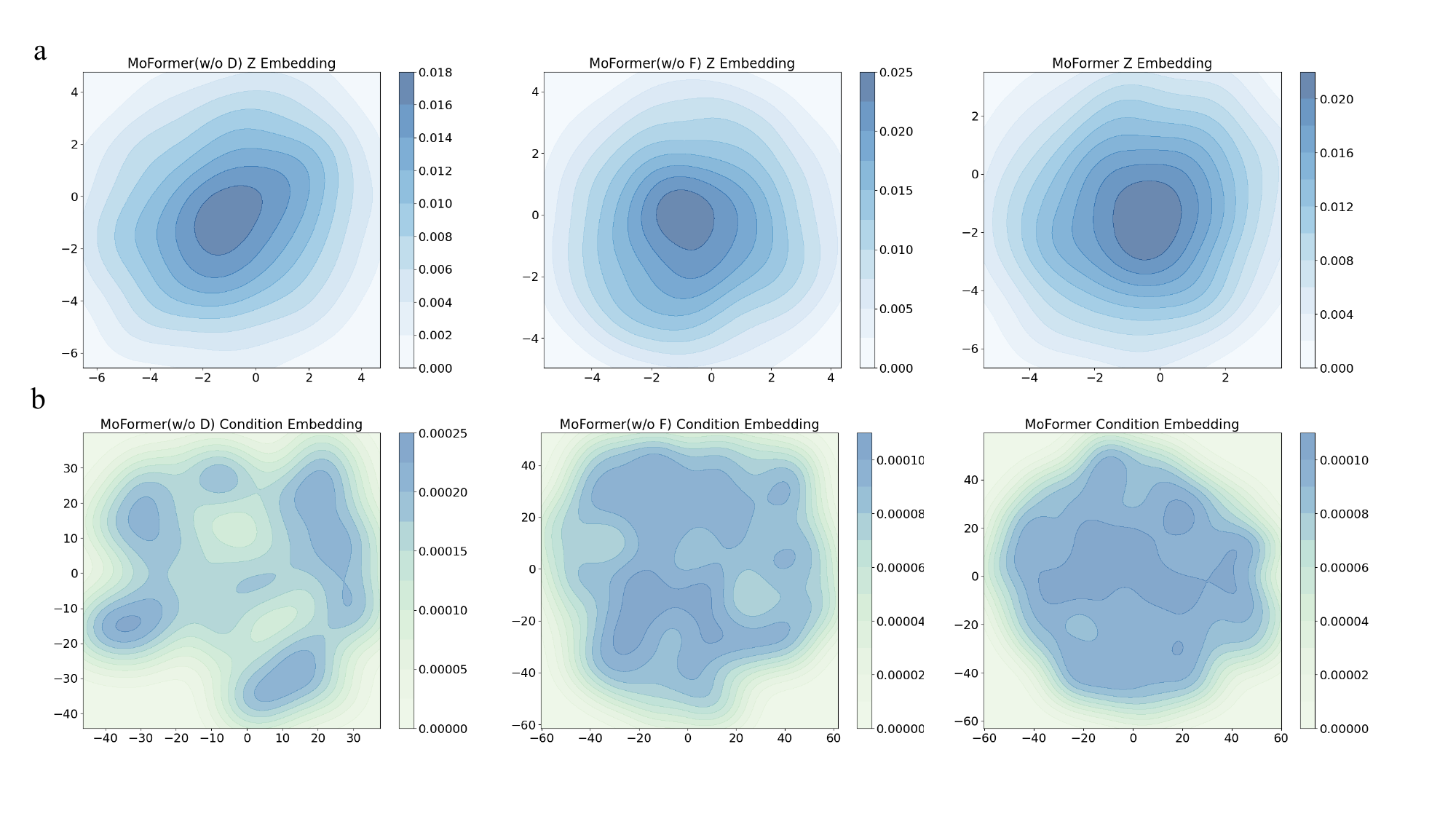}
\caption{Visualization of the encoding space $Z$ and condition landscape for MoFomer(w/o D), MoFomer(w/o F) and MoFomer.}
\label{latent_fig}
\end{figure}

We want to realize the optimization of multi-objective over a high dimensional discrete object space. As previously mentioned, we train a transformer, with encoders and decoders, to learn a continuous lower-dimensional embedding of multi-objection. The optimization is then performed in latent space and the best candidates are subsequently decoded into the original space. To gain a better understanding of how MoFormer optimize AMPs with respect to the attribute constraints and objectives, we provide visual illustration of the attribute landscapes and AMP embedding space as show in Figure \ref{latent_fig}. Specially, we leverage t-distributed Stochastic Neighbor Embedding (t-SNE) dimensionality reduction technique and Kernel Density Estimation (KDE) statistic skill to show our three well-designed methods (MoFomer(w/o D), MoFomer(w/o F), MoFomer), t-SNE effectively measures the similarity among high-dimensional data points, preserving these proximities when mapping to a reduced dimensional space, thereby facilitating dimensionality reduction; KDE estimates the overall data density by placing a kernel, such as a Gaussian, around each data point and aggregating the contributions from all kernels, which is particularly adept at revealing patterns and structures within the AMPs.

As shown in Figure \ref{latent_fig}(a), we can observe a two-dimensional (2D) density distribution of $z$ latent space, with the highest density in the central area, indicating that most data points are concentrated in this region. The intensity of the color correlates with the level of density, with darker hues signifying greater density. Observing the aggregation of data points, MoFomer(w/o F) exhibits a higher maximum density value compared to MoFomer(w/o D), indicating a more centralized data distribution. In addition, it can be noted that the maximum density value in MoFomer is lower than that in MoFomer(w/o D) and MoFomer(w/o F), suggesting a more dispersed data distribution and less concentration. This indicates that under the influence of all components, the model generates a more scattered feature representation, possibly signifying a richer expression in the latent space. The comparative analysis of these three methods affords us a deeper insight into the influence exerted by different model components on the distribution within the feature space. This observation aligns well with our intention of achieving a greater spread of features in the latent space, allowing for a more detailed capture of subtleties during the optimization process, thereby establishing the comprehensive MoFormer model as the preferable option.

Moreover, in figure \ref{latent_fig}(b), we can observe the latent space condition embeddings across three distinct configurations. For MoFomer(w/o D), the dispersion of data points across multiple high-density regions potentially reflects the model's heightened sensitivity to the nuances of data feature capture. In contrast, MoFomer(w/o F) exhibits a tendency for feature aggregation, with more pronounced and centralized high-density areas, especially in the core region, underscoring the pivotal role that three descriptors components plays in the orchestration of feature distribution. Lastly, configuration MoFomer is characterized by a focal convergence of high-density zones, culminating in a distribution that manifests as more homogenous, thereby facilitating a representation that is inherently more balanced. To sum up, the MoFormer framework adeptly captures the intricate distinctions in AMPs, achieving a harmonious and even feature distribution through the synergy of three descriptor operators and a multi-objective feedback strategy.

\section*{Discussion}
Recent work has impressively demonstrated the power of deep learning methods in discovering or design novel AMPs. However, these efforts still suffer from the contradictory and idealism-realism peptides that can be synthesized an tested, and the relatively long time it takes from design to validation. Here, we developed MoFormer to dramatically enhance the possiblility of multi-attribute balance in AMP design.

Compared to SOTA approaches, our success rate in finding multi-attributes is at a higher level, approximately doubles the performance of the least effective method. However, the supplementary experiments of AMP MIC and toxicity fail to manifest a commensurate level of marked superiority. This limitation in performance can potentially be ascribed to the paucity of toxicity benchmark data, impeding the efficacy of both the classifier and the generator. While recent advancements in deep learning approaches have been dedicated to studying AMP toxicity, yielding notable progress, the construction of a toxicity standard dataset still requires explicit modeling and calibration to address issues of data noise and scarcity.

In this study, we design comprehensive screening pipeline, a unique advantage of MoFormer that can execute a hierarchcal ranking of candidates, simultaneously employing multi-objective optimization algorithms alongside proxies fine-tuned on large-scale models. Each generator is capable of producing thousands of sequences of novel AMP candidates, yet only a limited number of these candidates can undergo experimental testing, a limitation that is intrinsically linked to the complex structure and properties of the candidates themselves. Under such circumstances, achieving candidate ranking through purely computational approaches can result in false positives or illusory advantages. Consequently, the development of a systematic, specialized, and human-computer interactive ranking methodology holds significant clinical implications.

While being unique and diverse, our de novo AMPs share common properties with known AMPs. They are predicted to be mostly $alpha$-helical peptides rich in cationic and hydrophobic amino acids and preferably act on negatively-charged IMs, showing that our pipeline was able to design new-to-nature sequences that follow the general building principles of AMPs. Overall, our work provides a proof-of-principle, how MoFormer can achieve multi-objective attribute optimization in AMP design. We describe the optimization and explanation of 11 de novo AMPs, of which all show multi-objective attribute advantages. Especially given the notably low clinical approval rates for AMPs, our approach heralds a novel paradigm in the quest for designing AMPs that are not only effective but also harmonize multiple attributes, thereby yielding a high degree of utility. This breakthrough offers a beacon of hope for the future development of AMPs.

\section*{Methods}

Language models are trained in a self-supervised fashion, a paradigm that uses the unstructured data itself to generate labels (e.g. the next word or the masked word in language models), and this paradigm possesses the capability to extract correlation across the whole sequences, transformer-based models achieve state-of-the-art (SOTA) performance in a variety of important tasks such as machine translation and question answering\cite{liu2019roberta,zhang2023applications}. Due to the notable parallels between the structure of human language and bioinformatics sequences, transformer-based approaches are increasingly recognized as highly effective for solving sequence-related issues within the field of bioinformatics.

\subsection*{Datasets}
For training samples, sequences with a length of up to 60 amino acids are collected from the studies by Witten et al\cite{witten2019deep}, Szymczak et al.\cite{szymczak2023discovering} and Hasan et al\cite{hasan2020hlppred}. After conducting multiple sequence alignment and applying a truncation ratio of 0.35 to remove redundancy, the final datasets now contain 4096 AMP sequences. These sequences are experimentally verified to have low MICs against E. coli, ranging from [-1,4], and exhibit a probability of hemolysis below 0.7. Additionally, to train a reliable proxy model, we collected 4547 samples of AMPs with antimicrobial activity ranging from [-1,4] from studies by Witten et al.\cite{witten2019deep} and Szymczak et al.\cite{szymczak2023discovering}, 1006 balanced hemolytic and non-hemolytic samples from the study by Hasan et al.\cite{hasan2020hlppred}, and 201 toxic samples from the study by Das et al.\cite{das2021accelerated} An equal number of non-toxic samples were randomly selected as controls. Subsequently, regression and classification models were fine-tuned based on the large Prot-BERT-BFD\cite{brandes2022proteinbert} model.

\subsection*{Sequence Modeling}
Given example sequences of the AMP $x=(x_1,…,x_n)$ composed from a finite alphabet of amino acids, the goal of language modeling is to learn $p(x)$. Because $x$ is a sequence, it is natural to factorize this distribution using the chain rule of probability:
\begin{equation}
p(x)=\prod_{i=1}^np(x_i|x_{<i})
\end{equation}

This process distills language modeling down to the prediction of amino acid sequences. Current SOTA methods train a neural network with parameters $θ$ to minimize the negative log-likelihood over a dataset $X=\{x^1,...,x^n\}{:}$:

\begin{equation}
\mathcal{L}(X)=-\sum_{k=1}^n\log p_\theta(x_i^k|x_{<i}^k)
\end{equation}

Given that language models are adept at learning the probability $p_\theta(x_i|x_{<i})$ a new AMP $x^{*}$ sequence with a specified length m can be synthesized by successively sampling its amino acids: $p_\theta(x_0),p_\theta(x_1|x_0^*),...,p_\theta(x_m|x_{<m}^*)$.

\subsection*{Sequence Modeling with MOMF}
Following the vanilla transformer, Zhao et.al.\cite{zhao2019variational} designed a method by formulating a conditional distribution $p(x|c)$, which can capture a property-specific prior on latent
representations. We are able to disentangle the target property $c$ from the prior through the construction of $p(z|c)$. Here, altering $c$ induces a change in the mean of the resultant Gaussian distribution. Therefore, continuous targets can be encoded in the mean of the distribution. For our study where target $c$ is constructed by multi-objective (MO) and multi-fused (MF) techniques for each AMP input $x$, the distribution can still be decomposed using the chain rule of probability and trained with a loss that takes the control code into account.
\begin{equation}
p(x|c)=\prod_{i=1}^np(x_i|x_{<i},c)
\end{equation}
\begin{equation}
\mathcal{L}(X)=-\sum_{k=1}^n\log p_\theta(x_i^k|x_{<i}^k,c^k)
\end{equation}
We formulate multi-objective constraints generation method by considering the conflicting attributes of AMPs. A multi-objective optimization problem (MOP) can be stated as: $C(x)=(c_1(x),...,c_j(x))^T$, subject to $x\in\Omega$ , where $\Omega$
is the decision space, $C:\Omega\to\mathbb{R}^j$ consists of $j$ real-valued objective functions and $\mathbb{R}^j$ is called the objective space. The attainable objective set is defined as the set $\{C(x)|x\in\Omega\}$. Very often, since the objectives in $C(x)$ contradict each other, no point in $\Omega$ maximizes all the objectives simultaneously. Take AMP two objectives generation as example, achieving potent antimicrobial activity (calculated by minimum inhibitory concentration, MIC) while avoiding hemolytic effects, a balance crucial for their safe therapeutic application. The corresponding objective function of this set-up takes the form:
\begin{equation}
\max\mathrm{imize~}C(x)=\left(c_{mic}(x),c_{hemo}(x)\right)
\end{equation}
where $(c_{mic}(x),c_{hemo}(x))$ represents the pair of constraints, mean square error is utilized for measuring the error: 

\begin{equation}
\mathcal{L}_{constraints}(X)=\sum_{k=1}^n\left.\left\{\left(c_{mic}(x^k),c_{hemo}(x^k)\right)-\left(c_{mic}^{\prime}(x^k),c_{hemo}^{\prime}(x^k)\right)\right\}^2\right. 
\end{equation}

In addition, we achieve intuitive techniques to force to model approaching low MIC and unlikely hemolytic. We define:
\begin{equation}
\begin{aligned}\mathcal{L}_{constraints}(X)&=\sum_{k=1}^n\left\{\left(c_{mic}(x^k),c_{hemo}(x^k)\right)-\left(c_{mic}^{\prime}(x^k),c_{hemo}^{\prime}(x^k)\right)\right\}^2\\\\&+\sum_{k=1}^n\left\{\left(c_{mic}^{\prime}(x^k)-\varepsilon\right)^2+\max\left(0,c_{hemo}^{\prime}(x^k)-\gamma\right)\right\}^2\end{aligned}
\end{equation}
where $\mathbf{\varepsilon}$ signify minimum value of MIC, empirically set at -2 (take log value). $\gamma$ serves as the input to the ReLU activation function, with 0.5 acting as the threshold. 

Although generative models have achieved promising results in conditional properties of AMPs, seeking a comprehensive and effective condition as the input for neural networks remains a challenge. To address this issue, we propose a multi-fusion (MF) AMP conditioning approach that leverages rich information beyond the original sequence, including physical, chemical, and structural aspects. Amino Acid Composition (AAC)\cite{roy2009exploiting} is a method utilized to delineate the frequency of amino acid occurrences within peptide sequences. This approach yields feature vectors characterized by multi-scale and continuous dimensions, with rapid transformation times. CTDC encompasses an analysis of peptide composition, transformation, and distribution\cite{chen2018ifeature}, indirectly offering insights into structural attributes. On the other hand, AAindex\cite{kawashima2000aaindex} delineates the physicochemical properties of amino acids, which in turn aids in representing peptide sequences effectively and enhances the discrimination between diverse peptide sequences within models. This tripartite fusion method takes into account the comprehensive representation of AMP functions, thereby generating a detailed and accurate 59-dimensional conditional description. Utilizing Principal Component Analysis (PCA)\cite{mackiewicz1993principal} for dimensionality reduction of conditional descriptors, we identify the most significant variances within the data to capture its essential features, represented in three dimensions. Subsequently, we integrate these features with attribute conditions to derive the final 5-dimensional conditional vector. Overall, conditional transformer can constraint posterior distribution by designing AMP with specific target attributes and comprehensive information. 

MoFormer learn $p_\theta(x_i|x_{<i},c)$ by training on AMP sequences prepended with control conditions. After elementary preprocessing, each AMP sequence, comprising $n$ tokens, is embedded as a sequence of $n$ corresponding vectors. Each vector is the aggregate of a learned token embedding and a sinusoidal positional embedding as in the original transformer architecture. The first core of the transformer block is the multi-head attention with $k$ heads that employs a causal masking strategy to prevent attending to future tokens:
\begin{equation}
\operatorname{Attention}(Q,K,V)=\operatorname{Softmax}\left(\frac{mask(QK^T)}{\sqrt{d}}\right)V
\end{equation}
where $d$ is the dimension of $K$, and $d$ must correspond to the dimension of $Q$.
\begin{equation}
\begin{aligned}
\operatorname{\text{Multi-He}ad}({Q},{K},V)& =\text{Concat(head}_1,\ldots,\text{head}_h)W_o ,  & 
  \mathrm{where~head}_i=\text{Attention}({Q}W_i^{Q},{K}W_i^K,VW_i^V) 
\end{aligned}
\end{equation}
where is the number of multi-heads, each $\mathsf{head}_i$ calculates an attention score between $Q$ and $K$ from different viewpoints using the different weights $(W_i^{Q},W_i^K,W_i^V)$ belonging to each head. The second core block is the a feedforward network with ReLU activation, parameters $W^{F}$ :
\begin{equation}
FF(X)=\max(0,XU)W^F
\end{equation}
Each block in the architecture commences with layer normalization, and follows it with a residual connection:
\begin{equation}
\begin{array}{ccc}\text{Block1:}&\bar{X}_i=\text{LayerNorm}(X_i)&H_i=\text{Multi-Head}(\bar{X}_i)+\bar{X}_i\\\text{Block2:}&\tilde{H}_i=\text{LayerNorm }(H_i)&X_{i+1}=\text{FF}(\overline{H}_i)+\overline{H}_i\end{array}
\end{equation}
During training, $X_{output}=(X_1,...,X_{i+1})$ is the input of a cross-entropy loss function: 
\begin{equation}
\mathcal{L}_{CE}(X)=CossEntropyLoss(X_{output},X^{\prime})
\end{equation}

\subsection*{Contrastive Regularization Strategy}
Despite the success of transformer, a dominant issue named posterior collapse has been observed, which greatly reduces the capacity of the generative model\cite{lucas2019don,takida2021preventing,zhao2019infovae}. The posterior collapse is usually formulated $D_{KL}[q_{\phi}(z|x)||p(z|c)]\to0$ , as for every $x$, and usually occurs when the decoder model is too powerful. This indicates that the learned variational distribution is almost identical to the prior (i.e., standard Gaussian distribution), resulting in the latent variables of different inputs indistinguishable in the latent space. To mitigate this issue, recent approaches have focused on diminishing the influence of the KL\text{-}divergence term  either by reducing its weight or by incorporating an extra regularization term. As a remedy, we introduced Kullback\text{-}Leibler divergence(KL)\cite{bowman2015generating} annealing in the loss function that is applied to stabilize the gradient flow and adopted dropout regularization strategy by a special gaussian dropout.
\begin{equation}
\mathcal{L}_{KL}(X)=\sum_{k=1}^n\delta D_{\mathrm{KL}}(q_\phi(z^k|x^k)||p_\theta(z^k|c^k))
\end{equation}
\begin{equation}
z^*=\mu+\alpha*\sigma\Theta\epsilon 
\end{equation}
\begin{equation}
\mathcal{L}_{GD}(X)=\sum_{k=1}^nCossEntropyLoss(z^k,z^{k^*})
\end{equation}
where ${\mathbf{\delta}}$ represents the weight of $D_\mathrm{KL}$, controling the gradient size, and $\alpha$ is the learnable weight parameter to control the gaussian dropout ratio. Blum et al.\cite{kingma2015variational} further interpreted $\alpha$ from bayesian regularization perspective, where the KL divergence of the posterior distribution and the prior is minimized, if an additional regularization term of $\alpha$ is jointly optimized with the model. We use the simple experience value 0.2 to set $\alpha$. In summary, plugging the loss illustrated above, we get the final objective function:
\begin{equation}
\mathcal{L}=\mathcal{L}_{CE}(X)+\mathcal{L}_{constraints}(X)+\mathcal{L}_{KL}(X)+\mathcal{L}_{GD}(X)
\end{equation}
where $\mathcal{L}_{CE}(X)$, $\mathcal{L}_{constraints}(X)$ represent reconstruction loss respectively, $\mathcal{L}_{KL}(X)$ signify Kullback-Leibler Divergence, and $\mathcal{L}_{GD}(X)$ is dropout regularization strategy.

\subsection*{MIC Regression and Hemolysis Classification}
In order to predict the properties of generated peptides, we trained a pair of classifiers that, for a given AMP sequence, predict its probabilities of being hemolytic and MIC active value against E. coli. Despite the abundance of predictors for MIC activity, the majority rely on a simplistic threshold-based classification of activity and inactivity, few methods have been developed to tailor a regression-based predictor for MIC active. In this work, we understand AMPs as all the peptides that are deposited in databases such as APD, CAMP, LAMP, and DBAASP. The exploration between MIC active and hemolytic is important as there exist multiple AMPs, which are not active against E. coli or are considered harmful to host cells. This motivates the use of a classifier and a regressor, where MIC value regressor classifier is trained on all the available AMP sequences, and MIC regressor is trained on a smaller subset of sequences with MIC measurements against E. coli available, and hemolytic classifier is trained on a limited number of sequences with hemolytic sequences. For the regression and classification of a potential AMP, we employ pretrained protein sequence model Prot-BERT-BFD\cite{brandes2022proteinbert}, is trained on a vast 2 billion protein fragments. Next, we use multiple fully connected layers and batch normalization layers to finetune model tailored for AMP attributes. The number of output units and the activation function of the final linear layer determine whether it performs regression for MIC values or classification for hemolytic and non-hemolytic properties. Both models are implemented in pytorch, resulting in extremely low mean squared error and surpasses 95\% accuracy on an independent validation set.

\subsection*{Pareto-based Feedback and Optimization}
In order to design  AMPs with desired pharmacological properties and increased diversity toward a biological target, we combine the pareto concept of a multi-objective optimization (MOP) algorithm with a feedback loop to train pooling and optimize the transformer-encoder. The constraint MoFormer generates a set of new AMPs, which are evaluated in terms of the MIC active against E. coli and hemolytic using the meticulously trained predictors. The multi-objective optimization algorithm sorts these AMPs based on the MIC active and hemolytic toward the biological pharmacological properties of AMP, where the top-ten PF rank candidates are used to feedback train pooling favoring the transformer-encoder. Hence, at each iteration step of this feedback optimization process, the transformer-encoder is fine-tuned with more desired properties by complementing a pareto-number of the input candidates with newly designed AMPs with preferable attributes.

The multi-objective optimization algorithm is pareto-based non-dominated sorting algorithm, which sets the solutions (AMPs) into multiple PFs according to the dominance across the objective functions (attributes). The PF incorporates solutions that are not inferior to any other solutions and are strictly superior in at least one objective function, i.e., it comprises non-dominated solutions. Therefore, the MOP seeks to identify a set of variables that optimizes two or more objective functions simultaneously. Considering a minimization problem, let $u$,$v$, $u$ is said to dominate $v$ if and only if  $u_i\geq\nu_i$ for every $\mathfrak{i}\in\{1,...,m\}\mathrm{~and~}u_j>v_j$ for at least one index $\text{j}\in\{1,...,m\}$.In the context of this work, the non-dominated sorting method is implemented using internal and external loops, where the former stores the solutions in distinct top PFs and the latter determines whether a given candidate dominates the solution space. The resulting optimal Pareto Front contains the AMPs with the tradeoff properties, i.e., lower values of MIC active, low risk of hemolysis. Overall, the proposed feedback optimization strategy of the MoFormer based on a multi-objective PF leads to the generation of candidates with desired properties, which to some extent addresses the dilemma of requiring specific attribute datasets.
 
\subsection*{Model Implementation Details}
MoFormer was implemented in PyTorch\cite{paszke2019pytorch} and trained using the Adam\cite{kingma2014adam} optimizer for gradient decent optimization, and the expansion rate of the momentum is 0.9 and the expansion rate of the adaptive term is 0.98, we apply warm-up algorithm to schedule the learning rate. WarmUp is intended to progressively raise the learning rate during the initial training phases, enhancing the model's ability to effectively explore the global optimal solution within the parameter space. The number of attention heads is set to 8, and each head dimension is updated for query, key and value via dividing the dimension of the input vector by the number of attention heads. We use the fully-connected layers with hidden dimension 512 and dropout rate 0.3 for predicting the initial sequence scores. We train our models for 25 epochs with a batch size 1024. All modeling experiments were carried out in about 2-2.5 hours on a single NVIDIA A100-PCIE-40GB.

\bibliography{sample}

\noindent LaTeX formats citations and references automatically using the bibliography records in your .bib file, which you can edit via the project menu. Use the cite command for an inline citation, e.g.  \cite{Hao:gidmaps:2014}.

For data citations of datasets uploaded to e.g. \emph{figshare}, please use the \verb|howpublished| option in the bib entry to specify the platform and the link, as in the \verb|Hao:gidmaps:2014| example in the sample bibliography file.

\section*{Acknowledgements (not compulsory)}

Acknowledgements should be brief, and should not include thanks to anonymous referees and editors, or effusive comments. Grant or contribution numbers may be acknowledged.

\section*{Author contributions statement}

Must include all authors, identified by initials, for example:
A.A. conceived the experiment(s),  A.A. and B.A. conducted the experiment(s), C.A. and D.A. analysed the results.  All authors reviewed the manuscript. 

\section*{Additional information}

To include, in this order: \textbf{Accession codes} (where applicable); \textbf{Competing interests} (mandatory statement). 

The corresponding author is responsible for submitting a \href{http://www.nature.com/srep/policies/index.html#competing}{competing interests statement} on behalf of all authors of the paper. This statement must be included in the submitted article file.

Figures and tables can be referenced in LaTeX using the ref command, e.g. Figure \ref{fig:stream} and Table \ref{tab:example}.

\end{document}